\documentclass[conference]{IEEEtran}

\usepackage[
  backend=biber,
  style=numeric,
  hyperref,
  maxnames=8,
  minnames=1,
  giveninits=true,
  maxcitenames=2,
  mincitenames=1,
  bibencoding=utf8,
  isbn=true,
  url=false,
  sorting=none
]{biblatex}
\usepackage[hidelinks]{hyperref}
\usepackage{tikz}
\usetikzlibrary{arrows.meta, positioning}
\usepackage{graphicx}
\usepackage{amsmath,amssymb,amsfonts}
\usepackage{mathtools}

\usepackage{algorithm}
\usepackage{algpseudocode}

\usepackage{tikz}
\usetikzlibrary{positioning,fit,arrows.meta,calc}

\usepackage{booktabs}
\usepackage{siunitx}
\usepackage{makecell}
\usepackage{comment}
\usepackage{xcolor}

\newtheorem{example}{Example}

\definecolor{ArrowIntra}{HTML}{6F1F00}
\definecolor{ArrowIntraTile}{HTML}{3333FF}
\definecolor{ArrowInterTile}{HTML}{009999}

\definecolor{MatchBlue}{RGB}{222,239,255}
\definecolor{MapeLow}{RGB}{255,235,238}
\definecolor{MapeMidLow}{RGB}{255,215,220}
\definecolor{MapeMed}{RGB}{255,205,210}
\definecolor{MapeHigh}{RGB}{239,154,154}

\begin{document}
\title{\huge Symbolic Polyhedral-Based Energy Analysis for \\ Nested Loop Programs}

\author{
  \IEEEauthorblockN{Avinash Mahesh Nirmala, Dominik Walter, Frank Hannig, and J\"urgen Teich}
  \IEEEauthorblockA{\{avinash.avi.mahesh, dominik.l.walter, frank.hannig, juergen.teich\}@fau.de}
  \IEEEauthorblockA{Hardware/Software Co-Design, Department of Computer Science,\\
  Friedrich-Alexander-Universit\"at Erlangen-N\"urnberg (FAU), Germany}
}
\maketitle

\begin{abstract}
This work presents a symbolic approach for estimating the energy consumption for nested loop programs when mapped and scheduled on parallel processor array accelerator architectures.
Instead of simulation-based evaluation, we derive a methodology for symbolic energy analysis that captures the impact of mapping and scheduling decisions of loop nests on  processor arrays.
We compare our approach against simulation-based results for selected benchmarks and varying sizes of the iteration spaces. Whereas the latter are not scalable, our symbolic analysis is shown to be independent of the problem size.
The presented evaluation methodology can be beneficially used during the design space exploration of mapping and scheduling decisions, for studying the influence of array size variations, and for comparisons with other loop nest accelerator architectures.

\end{abstract}

\IEEEpeerreviewmaketitle

\section{Introduction}\label{sec:Introduction}
There is a rapid growth of AI workloads in general.
But, particularly for applications running on embedded and edge platforms, power and energy budgets are very challenging to meet, especially as computational requirements grow.
Thereby, understanding and optimizing the impact of operations as well as data movements through a processing architecture is of utmost importance.
Often, the above workloads can be described adequately by nested programs for which dedicated accelerator architectures have been proposed, known as massively parallel processor architectures, including CGRAs, TCPAs \cite{TCPA1,TECS14,alpaka}, and GPUs.
For these, efficient compilation techniques exist to map loop nests to maximize performance.
However, missing is a highly accurate, yet efficient analysis of the impact of mapping decisions on energy consumption.
Whereas measurement-based approaches, being highly accurate, would require a physical prototype, simulation-based approaches at the register-transfer level (RTL) might be too time-consuming and thus not applicable for an early-stage energy analysis, such as during a design space exploration (DSE).
Moreover, the effect of tiling, scheduling, and mapping transformations on energy consumption has not yet been adequately studied for modern loop-intensive workloads. Obviously, accuracy and efficiency of evaluation are necessary, but contradictory
requirements for suitable analysis techniques.

Based on the above requirements, this paper presents an analytical framework for symbolic (parametric) evaluation of energy consumption of loop-intensive workloads when mapped and scheduled on processor array accelerator architectures.
The framework is applicable to a broad class of accelerator architectures and can be used at early design stages, thus neither requiring a hardware prototype nor time-intensive simulations.
    
The presented analysis methodology takes a loop nest with parametric loop bounds and a space-time mapping to a target processor array of given size. Using a one-time classification of energy costs for different memory and register accesses, we show how parametric expressions can be derived for the number of accesses and operations during the execution of a loop nest without actually executing it. These volumes of accesses and operations can be obtained symbolically, i.e., for loop nests whose bounds remain parametric. By combining these symbolic counts with architecture-specific per-access energy weights, our framework enables fast, fully analytical energy estimation without requiring cycle-accurate simulation or physical prototype measurements. Whereas the approach can be applied to a broad class of parallel processor architectures, we present an evaluation study focusing on Tightly Coupled Processor Arrays (TCPAs)~\cite{TCPA1,TECS14,alpaka}, which are massively parallel loop accelerator architectures that execute loop-intensive applications across grids of lightweight processing elements (PEs).

This paper is structured as follows: In Section~\ref{sec:Related Work}, we provide a review of existing work on energy estimation for massively parallel processor architectures.
In Section~\ref{sec:TCPA}, we introduce a class of VLSI processor arrays called TCPA, define a notation for loop nests, and present the basic methodologies for mapping and scheduling loop nests to such architectures.
In Section~\ref{sec:Symbolic_Loop_Nest_Energy_Analysis}, we then describe our symbolic energy analysis approach that is based on the efficient computation of volumes of polyhedral spaces of different types of memory access for which typical amounts of energy can be pre-characterized per access, thus providing a memory-centric evaluation of the overall energy consumption of a loop nest.
We show that such volume computations can be carried out even once only for parametric loop bounds, thus symbolically.
Section~\ref{sec:Experimental_results} gives evidence of the accuracy and speedup of this symbolic approach over a simulation-based approach.
Finally, Section~\ref{sec:Conclusion} summarizes and concludes the paper. 

\section{Related Work}\label{sec:Related Work}
Prior work on accelerator energy estimation largely assumes that execution behavior is fixed and known.
Activity-based analytical models, such as AccelWattch~\cite{AccelWattch}, estimate energy by weighting cycle-level activity statistics obtained from detailed simulation or hardware counters. Trace-driven architectural models, including CGRA-EAM~\cite{CGRA-EAM}, estimate energy by evaluating execution traces over pre-characterized functional units and interconnect components.
Component-level frameworks such as Accelergy~\cite{Accelergy} compute energy by associating architectural components with per-action energy costs and combining them with externally supplied action counts generated by analytical models, simulators, or mapping tools.

While effective for analyzing a fixed design and workload, these approaches share a common limitation: memory-access counts are evaluated only for explicitly specified workload instances. Even architecture-specific studies such as Eyeriss~\cite{Eyeriss} and analytical frameworks such as Timeloop~\cite{Timeloop} derive access counts only after fixing loop bounds, tensor dimensions, and a concrete dataflow, tiling, or mapping. Consequently, when workload sizes or execution order changes, the analysis must be recomputed.

In the polyhedral compiler community, \citeauthor{Verdoolaege04}~\cite{Verdoolaege04} launched the integer set library (ISL), which efficiently computes the volume of parametric polyhedra using Ehrhart polynomials, based on Barvinok's algorithm \cite{barvinok1994polynomial}.
In \cite{cache_analysis_using_barvinoks}, \citeauthor{cache_analysis_using_barvinoks} apply this calculus for the analysis of cache effects on execution time in single-core architectures.
    
In this paper, we rather analyze the energy consumption of loop nests when executed on massively parallel processor arrays in which many tiny processing elements do not even carry a cache.
For these, we show how to derive closed-form expressions relating tiling choices, execution schedules, and resource bindings to energy consumption, including operations and memory accesses.
This calculus enables an efficient analytical evaluation of energy consumption that is parametric in the problem size (loop bounds). To the best of our knowledge, this is among the first approaches to symbolically analyze the energy consumption of nested-loop programs in dependence on space–time mapping decisions.
    
In the domain of processor array accelerators, coarse-grained reconfigurable arrays (CGRAs)~\cite{26_CGRA_Overview,24_CGRA_Taxonomy} represent a prominent class of architectures that employ an \emph{operation-centric} mapping approach~\cite{cgra_comp}, where the operations from a data flow graph (DFG) are individually assigned to processing elements.
For these, simulation is usually used to analyze performance and energy tradeoffs of a compiled loop nest. But obviously, this is not a scalable approach. In order to exploit  scalable symbolic energy analysis techniques as introduced in this paper, a polyhedral iteration space representation is necessary.
Whereas CGRA compilers usually operate at the granularity of individual operations of a single-dimensional loop, other classes of processor array architectures, such as tightly coupled processor arrays (TCPAs) \cite{TCPA1, TECS14}, start 
the compilation from a given polyhedral representation of a loop nest to determine schedules and mappings over tiled, multidimensional iteration spaces~(\emph{iteration-centric} mapping).
Thus, although in general the symbolic approach described in the following could also be applied to many CGRA architectures from a hardware perspective, compilation approaches would only benefit when using parametric polyhedral loop descriptions.     
    
\section{Loop Nests and their Mapping and Scheduling on Processor Arrays}\label{sec:TCPA}
This section introduces a class of massively parallel processor arrays known as TCPAs \cite{TCPA1, TECS14}, and outlines the core concepts of loop-nest representation, space–time mapping, and symbolic loop scheduling \cite{TeichTH4}.

\subsection{Tightly Coupled Processor Arrays (TCPAs)}\label{sec:TCPAarchitecture}
TCPAs \cite{TCPA1, TECS14} are parallel processor array architectures designed to execute multidimensional, loop applications represented as polyhedral recurrence equations as piecewise regular algorithms (PRAs). A TCPA works with a host, such as a CPU or FPGA. In this section, we provide an overview of the TCPA architecture.
Next, we introduce the PRA representation for expressing loop nests and explain how iteration spaces are partitioned, scheduled, and mapped for spatial execution.
        
        \label{sec:TCPA_arch}
            \begin{figure}[h]
                \centering
            	\includegraphics[width=0.37\textwidth]{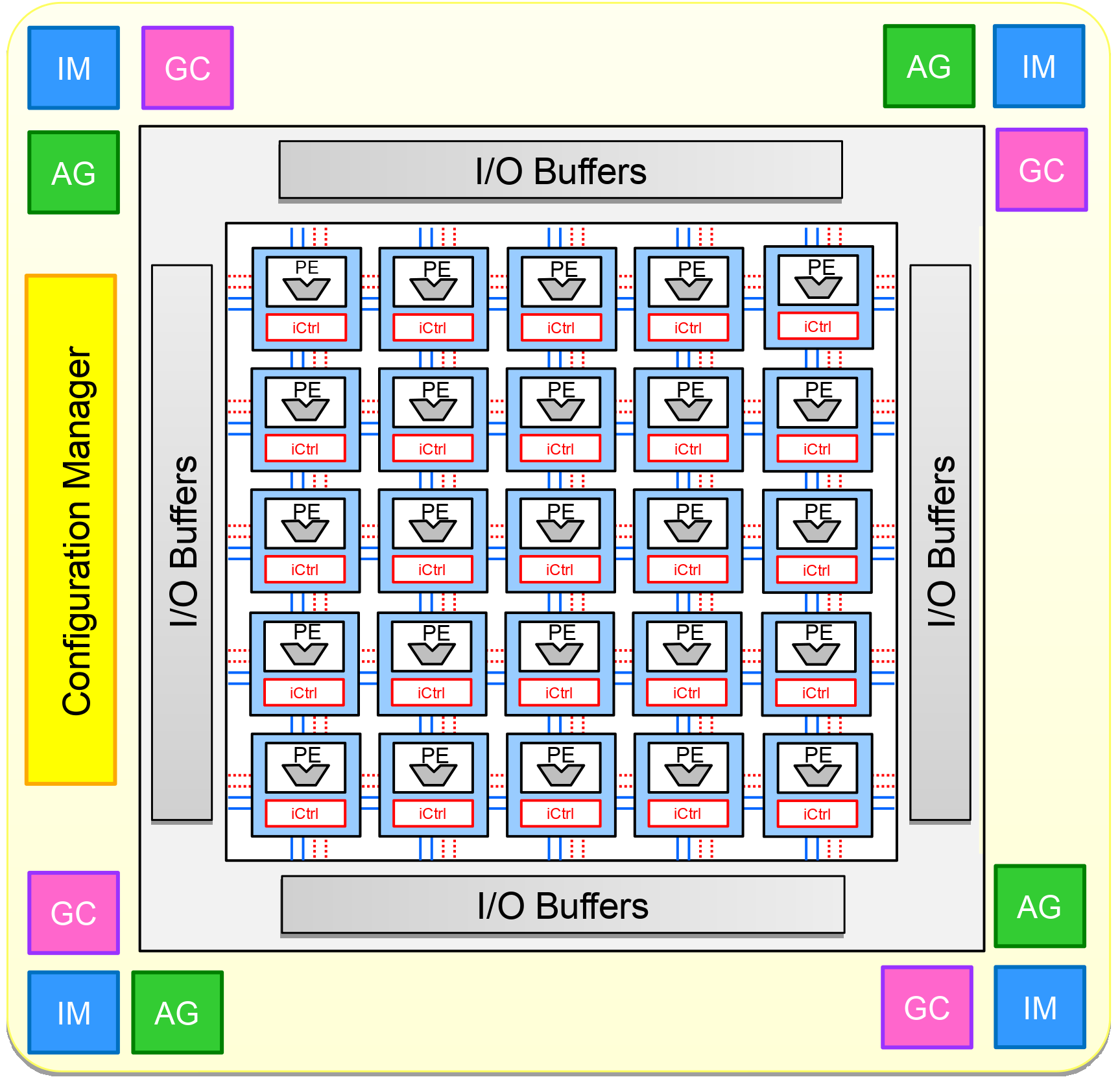}
            	\caption{Example of a tightly coupled processor array (TCPA) architecture~\cite{TCPA1}.}
            	\label{fig:tcpa_arch}
        \end{figure}

        TCPAs are two-dimensional arrays of programmable processing elements (PEs) connected by a configurable circuit-switched neighbor-to-neighbor interconnect, as shown in Figure \ref{fig:tcpa_arch}. The PE grid is bordered by four input/output (I/O) buffers, each comprising multiple dedicated address generators (AG). The architecture also includes peripheral control units, a global controller (GC), and a configuration manager (CM), which orchestrates loop-program execution on the array. The TCPA operates as a memory-coupled coprocessor alongside a host system. The loop I/O controller schedules direct memory access (DMA) to transfer data between the host's external memory and the I/O buffers. After the TCPA is configured by the host, it executes the loop program independently, while the loop I/O controller schedules DMA transfers during execution to support dynamic refilling.
        
        TCPAs avoid costly direct PE-to-DRAM communication by transferring data between the host and I/O buffers rather than having the PEs access them directly. During execution, all active data movement is restricted to the I/O buffers and the PE array \cite{IO_transfor}. This ensures predictable, high-bandwidth on-chip communication.

        Each PE can be configured with multiple parallel functional units (FUs). Each FU has its own instruction memory, branch unit, and program counter. Input and output data streams between the I/O buffers and the PEs, while all intermediate values remain within the PE's internal register hierarchy. Within each PE, the architecture supports multiple register classes, namely general-purpose, feedback, and input/output registers. The compiler~\cite{WitteraufWHT21} maps data to these register classes based on dependence structures and schedules. During execution, values produced and consumed by operations are stored and forwarded through these registers according to architectural constraints, and the circuit-switched interconnect supports inter-PE communication when dependencies span multiple PEs, while the exact mapping rules are defined in the binding phase and discussed later.
    
\subsection{Loop-Nest Representation}\label{sec:PRA}
Almost any standard programming language supports loop nests syntactically, e.g., C, C++, Java, and Python.
        In this paper, we assume loop nests described using a polyhedral notation called \emph{Piecewise Linear Algorithms (PLAs)} \cite{TeichT93, Teich93}. Normal sequential for-loop specifications can be converted  to such a polyhedral description for parallelization.
A PLA describes an $n$-dimensional loop nest by an iteration space $\mathcal{I} \subseteq \mathbb{Z}^n$, where each element $\mathbf{i} = (i_0, i_1, \dots, i_{n-1})^{\mathrm{T}} \in \mathcal{I}$, called \emph{iteration vector}, corresponds to one loop iteration. In the following, we 
assume $\mathcal{I}$ is described by a polyhedral description $\mathcal{I} = \{\,\mathbf{i} \mid \mathbf{A} \mathbf{i} \ge \mathbf{b}\,\}$, where $A \in \mathbb{Z}^{m \times n},\; b \in \mathbb{Z}^n$.
        The computations of a loop nest are described by a set of quantified statements $S = \{\ldots, S_q \ldots\}$, with $S_q$ given by 
        \begin{equation}
        \label{eq:pla-statement}
        \begin{aligned}
        S_q:\;& x_q[\mathbf{P}_q\mathbf{i}+\mathbf{f}_q] 
             = F_q(\ldots, x_{q,r}[\mathbf{Q}_{q,r}\mathbf{i}-\mathbf{d}_{q,r}],\ldots)\; \text{if } \mathbf{i} \in \mathcal{I}_q.
        \end{aligned}
        \end{equation}
        For all $\mathbf{i} \in \mathcal{I} \cap \mathcal{I}_q$ (with $\mathcal{I}_q$ called {\em condition space}), the variable $x_q$ at index (vector) $\mathbf{P}_q \mathbf{i} + \mathbf{f}_q$ is defined as value of the function $F_q$ evaluated on variables $x_{q,r}$ at $\mathbf{Q}_{q,r}\mathbf{i} - \mathbf{d}_{q,r}$. We finally call a PLA \emph{Piecewise Regular Algorithm} (or PRA) for the special
        case that $\mathbf{P}_q$ and $\mathbf{Q}_{q,r}$ are identity matrices and $\mathbf{f_q}$ is zero. As a result, a statement of a PRA has the following form:
        \begin{equation}
        \label{eq:pra-statement}
        \begin{aligned}
        S_q:\;& x_q[\mathbf{i}] 
             = F_q(\ldots, x_{q,r}[\mathbf{i}-\mathbf{d}_{q,r}],\ldots)\; \text{if } \mathbf{i} \in \mathcal{I}_q.
        \end{aligned}
        \end{equation}
        Note that in a PLA or PRA, there exists neither an explicit order of execution of iterations like in imperative loop programs, nor any order of execution of statements.
        Instead, each quantified statement only implies implicit schedule restrictions by so-called {\em data dependencies}. For example, in a PRA with a statement given in Eq.~(\ref{eq:pra-statement}), the left-hand-side variable $x_q[\mathbf{i}]$ can only be evaluated
        after all variables $x_{q,r}[\mathbf{i}-\mathbf{d}_{q,r}]$ (arguments of $F_q$)
        on the right-hand side have been computed. We also say $x_q[\mathbf{i}]$
        depends on $x_{q,r}[\mathbf{i}-\mathbf{d}_{q,r}]$, and call the constant vector
        $\mathbf{d}_{q,r}$ the {\em dependence vector}.
        Finally, instances of variables appearing only on the right-hand side of statements are called \emph{input variables}. Similarly, instances of variables appearing only on the left-hand side of statements
        are called \emph{output variables}. All other instances of variables are called \emph{internal variables}.

        \begin{example}
        \label{ex_1}
        In the following, we introduce as a running example the nested loop program GESUMMV from the PolyBench~\cite{PolyBench} benchmark, which computes the sum of two matrix–vector products involving a vector $x \in \mathbb{Z}^{N_1}$ and matrices $A, B \in \mathbb{Z}^{N_0 \times N_1}$:
        \begin{equation*}
        Y[i_0] =
        \sum_{i_1=0}^{N_1-1}
        \bigl(A[i_0,i_1]\cdot X[i_1] + B[i_0,i_1]\cdot X[i_1]\bigr),
        \ \ 0 \le i_0 < N_0 .
        \end{equation*}
        A corresponding 2-dimensional PRA can be formulated with an iteration space $\mathcal{I} = \{ (i_0,i_1) \mid 0 \le i_0 < N_0,\; 0 \le i_1 < N_1 \}$ and the following set of statements:
        {\small
        \begin{equation*}
        \begin{aligned}
        S_{1}:~& x[i_0,i_1] = X[i_1] \qquad &&\text{if } i_0 = 0\\
        S_{2}:~& x[i_0,i_1] = x[i_0-1, i_1] \qquad &&\text{if } i_0 > 0\\
        S_{3}:~& a[i_0,i_1] = A[i_0,i_1]\cdot x[i_0,i_1]\\
        S_{4}:~& b[i_0,i_1] = B[i_0,i_1]\cdot x[i_0,i_1]\\
        S_{5}:~& s_A[i_0,i_1] = a[i_0,i_1] \qquad &&\text{if } i_1 = 0\\
        S_{6}:~& s_A[i_0,i_1] = s_A^*[i_0,i_1] + a[i_0,i_1] \qquad &&\text{if } i_1 > 0\\
        S_{7}:~& s_A^*[i_0,i_1] = s_A[i_0,i_1-1] \qquad &&\text{if } i_1 > 0\\
        S_{8}:~& s_B[i_0,i_1] = b[i_0,i_1] \qquad &&\text{if } i_1 = 0\\
        S_{9}:~& s_B[i_0,i_1] = s_B^*[i_0,i_1] + b[i_0,i_1] \qquad &&\text{if } i_1 > 0\\
        S_{10}:~& s_B^*[i_0,i_1] = s_B[i_0,i_1-1] \qquad &&\text{if } i_1 > 0\\
        S_{11}:~& Y[i_0] = s_A[i_0,i_1] + s_B[i_0,i_1] \qquad &&\text{if } i_1 = N_1 - 1
        \end{aligned}
        \end{equation*}
        }
        In the above description, all instances of variables $A$, $B$, and $X$ are input variables. Similarly, all instances of variable $Y$ are output variables.  
        Instances of $a$ and $b$ represent the element-wise products of $A \cdot X$ and $B \cdot X$. Instances $s_A$ and $s_B$ are partial sums accumulated along the dimension $i_1$ (statements $S_{6}, S_{7}$ and $S_{9},S_{10}$), initialized in statements $S_{5}, S_{8}$, respectively (at $i_1 = 0$). The outputs $Y[i_0]$ are obtained at $i_1 = N_1 - 1$ (statement~$S_{11}$).
    \end{example}
    
Now, in order to map a polyhedral loop nest to a 1- or 2-dimensional array of processing elements, the given iteration space $\mathcal{I}$ is partitioned into as many tiles as available processing elements.
For each tile, then a feasible set of iterations within a tile and between tiles needs to be found.
This is explained in the following.

    \subsection{Symbolic Tiling}\label{subsec:symbolic-tiling}
        According to~\cite{WitteraufWHT21, TeichT93, TeichTZ97}, we partition the $n$-dimensional iteration space $\mathcal I$ into congruent rectangular tiles $\mathcal J$ by using a partitioning 
        matrix $P = {\mathrm{diag}}(p_0, \ p_1, \ldots, p_{n-1})$ such that 
        $\mathcal I \subseteq \mathcal J \boldsymbol{\oplus} P \mathcal K$.
        A \emph{tile} $\mathcal J$ can be described as follows:
        \begin{equation}
        \mathcal J
        = \bigl\{\, \mathbf j = (j_0, \ldots, j_\ell, \ldots, j_{n -1})^\mathrm T \mid  0 \le j_\ell < p_\ell \,\bigr\}.
        \end{equation}
        Similarly, the \emph{set of} non-empty \emph{tile origins} $\mathcal K$  can be described as:
        \begin{equation}
        \mathcal K
        = \bigl\{\, \mathbf k = (k_0, \ldots, k_\ell, \ldots, k_{n -1})^\mathrm T \mid 0 \le k_\ell < t_\ell \,\bigr\}.
        \end{equation}
        The vector $T= (t_0, \ldots, t_\ell \ldots, t_{n-1})$ describes the number of tiles along each dimension~$\ell$.
        After tiling, the original dependencies in statements as shown in Eq.~(\ref{eq:pra-statement}) $\mathbf{d} \in \mathbb{Z}^n$ are decomposed into an {\em intra-tile dependence vector} ($\mathbf{d}_J$) and an {\em inter-tile dependence vector} ($\mathbf{d}_K$) and all instances of variables are embedded into the $2n$-dimensional space with dependence vector ($\mathbf{d}_J,\mathbf{d}_K)^\mathrm{T}$.
        Without loss of generality, we can split the statement $S_q$ in Eq.~(\ref{eq:pra-statement}) into two equations below \cite{TeichT93}:
        \begin{equation}
            S_q:\;
            x_q[\mathbf{j}, \mathbf{k}]
            = F_q(\ldots,\, x_{q,r}^{\ast}[\mathbf{j}, \mathbf{k}],\, \ldots)\; \text{if } \mathbf{j} + P \mathbf{k} \in \mathcal{I}_q
            \label{eq:displacement_1}
        \end{equation}
        \begin{align}
            S_q^{\ast}:\;
            x_{q,r}^{\ast}[\mathbf{j},\, \mathbf{k}]
            = & x_{q,r}[\mathbf{j}- d - P \gamma,\, \mathbf{k} + \gamma]\; \text{if } \mathbf{j} + P \mathbf{k} \in \mathcal{I}_q \;  \nonumber \\
            & \wedge \mathbf{j} - d - P \gamma \in \mathcal J.
            \label{eq:displacement_2}  
        \end{align}
        Whereas the dependencies of the transformed statement $S_q$ are all zero, we need to create one statement
        of the form $S_q^{\ast}$ for each solution $\gamma$ that satisfies the set of inequalities 
        \begin{equation}
        \label{eq:gamma}
        \{\gamma \in \mathbb{Z}^n : -e < \gamma + P^{-1} d < e \}
        \end{equation}
        and $e = (1 \; 1 \, \cdots 1)^{\mathrm T}.$


        \begin{figure}[h]
            \centering
        	\includegraphics[width=0.43\textwidth]{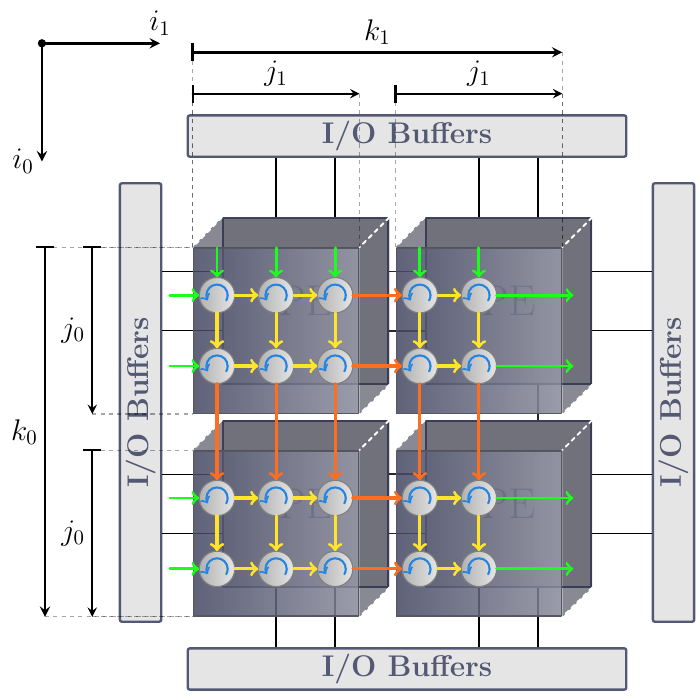}
        	\caption{Visualization of the tiled iteration space of the GESUMMV benchmark for an iteration space of size $N_0 \times N_1 = 4 \times 5 = 20$ on a processor array of size $t_0 \times t_1 = 2 \times 2 = 4$ PEs and tiles of size $p_0 \times p_1 = 2 \times 3$. The dependencies are also indicated by arrows. Green arrows denote I/O buffer accesses. 
            }
        	\label{fig:tcpa_arch_mapping}
        \end{figure}
    
    \begin{example}
    \label{ex_2}
        In the following, we tile the iteration space of the GESUMMV PRA listing
        introduced in Example~\ref{ex_1}. Assume a matrix of size
        $N_0 \times N_1 = 4 \times 5$ to be mapped onto a $2 \times 2$ target processor array. Figure~\ref{fig:tcpa_arch_mapping} provides a visual illustration of the
        tiling into congruent tiles
        $\mathcal J
        = \bigl\{\, \mathbf j = (j_0, j_1)^\mathrm T \mid  0 \le j_\ell < p_\ell \,\bigr\}$ of size $p_0 \times p_1 = 2 \times 3 = 6$ iterations.
        This tile size was chosen to partition the
        iteration space into exactly as many PEs as are available in each dimension of the processor
        array. Thus, the resulting set $\mathcal K$ of tile origins is
        $\mathcal K
        = \bigl\{\, \mathbf k = (k_0, k_1)^\mathrm T \mid 0 \le k_\ell < t_\ell \,\bigr\}$
        with $t_0 = t_1 =2$, thus a total of $t_0 \times t_1 = 2 \times 2 = 4$ tiles.
        For the explanation of the transformation of data dependencies due to tiling, consider statement $S_7$ as an example after tiling.
        According to Eq.~\eqref{eq:displacement_2},
        we obtain the following transformed statement $S_7^*$.
        \begin{align} 
        S_7^{\ast}:\;
        s_A^{\ast}[\mathbf{j}, \mathbf{k}]
        &=
        s_A[\mathbf{j}- (0, 1)^{\mathrm{T}} - P\gamma,\; \mathbf{k}+\gamma] \; 
        \text{if } j_1 + p_1 k_1 > 0 \nonumber
        \\
        &\wedge\ \mathbf j- (0, 1)^{\mathrm{T}}-P\gamma \in \mathcal J. \nonumber
        \end{align}
        for which we can find two solutions for the vector $\gamma$ that satisfy Eq.~(\ref{eq:gamma}):
        $\mathcal{\gamma} = \left\{(0,0)^{\mathrm T},\ (0,-1)^{\mathrm T}\right\}.$
        The final new equations for $S_7^*$ that replace $S_7^{\ast}$, are thus:
        \begin{align}
        S_{7}^{\ast 1}:\;
        s_A^{\ast}[\mathbf{j}, \mathbf{k}]
        &=
        s_A[\mathbf{j}- (0, 1)^{\mathrm{T}},\; \mathbf{k}] \; 
        \text{if } j_1 + p_1 k_1 > 0 \nonumber
        \\
        &\wedge\ \mathbf j- (0, 1)^{\mathrm{T}} \in \mathcal J. \nonumber
        \end{align}
        \begin{align}
        S_7^{\ast 2}:\; s_A^{\ast}[\mathbf{j}, \mathbf{k}] = s_A[\mathbf{j}- (0, 1-p_1)^{\mathrm{T}},\; \mathbf{k} + (0,-1)^{\mathrm{T}}] \; \nonumber 
        \\
        \text{if } j_1 + p_1 k_1 > 0 \nonumber
        \wedge\ \mathbf j- (0, 1-p_1)^{\mathrm{T}} \in \mathcal J. \nonumber
        \end{align}
        The two new resulting dependence vectors 
        \begin{equation*}
        \label{eq:dstar_s4}
        \mathbf d_6^{\ast}\!\bigl((0,1)^{\mathrm T}\bigr)
        =
        \left\{
        (0,1,0,0)^{\mathrm T},\;
        (0,1-p_1,0,1)^{\mathrm T}
        \right\}
        \end{equation*}
        are also visualized in Figure~\ref{fig:tcpa_arch_mapping}.
        The first vector (shown in yellow) corresponds to an intra-tile dependence
        and leads to an intra-processor memory access, whereas the second
        (shown in orange) represents an inter-tile dependence that leads to an inter-processor
        memory access and communication, as we exploit later in our energy analysis.
    \end{example}

  Now, in order to execute the iterations within each tile and to respect data dependencies both within a tile and across tiles, a feasible schedule must be determined. This is explained in the following.
   
    \subsection{Symbolic Scheduling}
    \label{subsec:symbolic-scheduling}
        Whereas tiling defines the mapping of iterations and their corresponding
        computations to PEs, we still need to find execution
        schedules that minimize execution time.
        A schedule assigns a variable $x_q$ defined in a statement $S_q$ a start time $t_q(J,K)$ when the operation $F_q$ is computed for each iteration $(J,K)$.
        According to~\cite{TeichTH4,WitteraufWHT21}, iterations within a tile must be executed
        in either a sequential or a pipelined (modulo-scheduled) order with an
        initiation interval $\pi$ between two iterations. Different tiles are scheduled in parallel across a PE array, such an execution corresponds to a {\em locally sequential, globally parallel (LSGP) modulo schedule}.
        Moreover, the regular computations of loop nests can be efficiently scheduled using linear schedules in which the
        schedule of iterations within a tile is described by an {\em intra-tile schedule vector} $\boldsymbol{\lambda}^{J}$. Similarly, the start time of tiles is also described by a so-called {\em inter-tile schedule vector} $\boldsymbol{\lambda}^{K}$.
        According to \cite{sym_tiling_sche}, schedules $(\boldsymbol{\lambda}^{J},\boldsymbol{\lambda}^{K})$ that minimize
        the {\em global latency} $L$ of a loop nest for a given initiation interval $\pi$ can be determined efficiently as follows:
        \begin{equation}
        \label{eq:global-latency}
        L \;=\;
        \boldsymbol{\lambda}^{J}\begin{pmatrix}p_0 - 1\\\vdots\\p_{n-1} -1\end{pmatrix} 
        \;+\;
        \boldsymbol{\lambda}^{K}\begin{pmatrix}t_0 - 1\\\vdots\\t_{n-1} -1\end{pmatrix}         
        \;+\;
        L_c.
        \end{equation}
        The first term determines the maximal difference of start times between any two tiles.
        The second term determines the maximum difference of any pair of iterations within a tile.
        Finally, $L_c = \max_{1 \le q \le |S|} (\tau_q + w_q)$ describes the latency of a single iteration within a tile, where $\tau_q$ is the start time and $w_q$ the execution latency of operation $F_q$ in statement $S_q$.
        
    \begin{example}
    \label{ex_3}
        Given the tiling of the iteration space of the GESUMMV program introduced in Example~\ref{ex_2}. For the given application, and assuming a pipeline
        interval $\pi = 1$, we can find
        the symbolic intra-tile schedule $\lambda^{J} = (1, p_0)^\mathrm{T}$
        and the inter-tile schedule $\lambda^{K}= (p_0, p_0 \cdot (p_1-1) +1)^\mathrm{T}$,
        minimizing the global latency given by
        $L \;=\;
        \lambda^{J}  \cdot (p_0-1, p_1-1)^\mathrm{T} + \lambda^{K} \cdot (t_0 - 1, t_1 - 1)^\mathrm{T}      
        \;+\;
        L_c$ with $L_c = 4$, assuming each $F_q$ has a latency of $w_q=1$, which leads to $L = (p_0 p_1 - 1) + p_0 (t_0 - 1) + (p_0 (p_1 - 1) + 1)(t_1 - 1) + 4$.
    In the running example with $p_0=2,p_1=3,t_0=t_1=2$, we obtain $L = 5 + 7 + 4 = 16$.
    \end{example}
      
\section{Symbolic Loop Nest Energy Analysis}\label{sec:Symbolic_Loop_Nest_Energy_Analysis}
    In this section, we introduce our symbolic energy analysis methodology for
    loop nests when executed on processor arrays.
    The methodology starts by analyzing both the {\em computational energy} (for execution of loop statements of type $S_q$ in Eq.~(\ref{eq:displacement_1})) and
    the {\em data transport energy} for any data movement, distinguishing both from outside the chip boundary, e.g., DRAM, and inside the processor array and back (by analyzing loop statements of type $S_q^{\ast}$ in Eq.~(\ref{eq:displacement_2})).
    This first part of the analysis delivers a formula for the energy
    consumption of each loop statement per loop iteration.
    Subsequently, these energy-by-statement expressions are multiplied
    by the number of iterations each statement is executed to deliver a final overall energy estimate. This step relies on the efficient calculation of volumes of polyhedral spaces related to the size of the condition space of each loop statement. 
We demonstrate that these volumes can be computed symbolically for parametric loop bounds, thus enabling the energy analysis to be computed only once for parametric loop bounds and ultra-fast scalability analysis by simply inserting a set of concrete loop bound values into the derived volume formulae.
    
    \subsection{Energy-by-Statement Analysis}
    \label{subsec:Binding}
    Without loss of generality, we can assume that any PRA loop statement
    as described in the most general form in Eq.~(\ref{eq:pla-statement}) can be split into one statement with only zero dependence vector right-hand side (RHS) arguments $S_q$ but containing computations expressed
    by a function ${\cal F}_q$ as described in Eq.~(\ref{eq:displacement_1}) and statements that just relate each RHS variable with variables at displacements,
    i.e., statements of type $S_q^*$ in Eq.~(\ref{eq:displacement_2}).
    This eases the description of the following energy analysis by splitting
    the analysis of {\em computational energy}
    and the analysis of {\em memory-related energy}. In the following,
    we denote $C$ the set of {\em computational statements},
    e.g., and let $T$ denote the set of {\em memory statements}.

    \begin{example}
    \label{ex_4}
    In the running example introduced in Ex.~\ref{ex_1}, all 11 statements already adhere
    to the above form with $C=\{S_3,S_4,S_6,S_9,S_{11}\}$ denoting
    statements with computations and $M= \{S_1,S_2,S_5,S_7,S_8,S_{10}\}$ being the set of memory-related statements.
    \end{example}

    A processor array, as shown in Figure~\ref{fig:tcpa_arch_mapping}, comprises a memory system consisting of 6 different memory types $\tau \in \mathcal{T}$
    as classified in the above table: on-chip I/O buffers ($\mathrm{IO}$) at the periphery,
    through which data transfers for fetching and storing tensor I/O data from/to a host DRAM ($\mathrm{DR}$)
    are performed explicitly via DMA operations.
Because DRAM accesses are the most expensive, the loop mapping strategy should avoid moving any input or output variable instance of a loop nest across the chip boundary more than once.
    Moreover, within each PE, we distinguish four classes of registers
    with the following use: \emph{general-purpose registers} ($\mathrm{RD}$) for handling intra-iteration dependencies,
    \emph{feedback registers} ($\mathrm{FD}$) for temporarily storing variables with inter-iteration dependencies that are processed locally within a PE, and finally \emph{input/output registers} ($\mathrm{ID}$ and $\mathrm{OD}$)
    used to receive data from, or transmit data to, neighboring PEs via communication ports.
    A breakdown of typical energies per access for different types of memory accesses ${\mathrm T}
    = \{\mathrm{RD},\mathrm{FD},\mathrm{ID},\mathrm{OD},\mathrm{IOb},\mathrm{DR}\}$ and for some basic arithmetic operations
     is given in Table~\ref{tab:energy_weights} for a \qty{45}{\nano\meter} technology~\cite{energy_weights}.

{\footnotesize
\begin{table}[t]
\centering
\caption{Energy related to memory accesses/operations in \qty{45}{\nano\meter} technology~\cite{energy_weights}.}
\label{tab:energy_weights}
\renewcommand{\arraystretch}{1.05}
\setlength{\tabcolsep}{4pt}
\begin{tabular}{|l|c|}
\hline
\multicolumn{1}{|c|}{\textbf{Memory Class/Operation Type}} & \textbf{Energy $E$ [\unit{\pico\joule}]} \\
\hline
\multicolumn{2}{|c|}{\textbf{Register Files}} \\
\hline
General-purpose register ($\mathrm{RD}$) & 0.12 \\
Feedback register ($\mathrm{FD}$)        & 0.35 \\
Input register ($\mathrm{ID}$)           & 0.24 \\
Output register ($\mathrm{OD}$)          & 0.12 \\
\hline
\multicolumn{2}{|c|}{\textbf{Buffers / Off-chip Access}} \\
\hline
I/O buffer ($\mathrm{IOb}$)                & 16 \\
DRAM    ($\mathrm{DR}$)             & 1280 \\
\hline
\multicolumn{2}{|c|}{\textbf{Arithmetic Operations}} \\
\hline
Addition ($\mathrm{add}$)       & 0.36 \\
Multiplication ($\mathrm{mul}$) & 1.24 \\
\hline
\end{tabular}
\end{table}
}

\subsubsection{Energy Computational Statements}
 We start with the analysis of energy related to computational statements $S_q \in C$.
 Let $L: x \rightarrow \mathrm{T}$ denote the function that returns the {\em memory class}
 of a variable $x$ occurring either on the LHS (write location) or the RHS (read locations).
 Then, the computational energy for one loop iteration of such a statement $S_q$ can be estimated by:
 \begin{equation}
   E_q^C = \sum_{r=1}^R E(L(x_{q,r}^*)) + E({\cal F}_q) + E(L(x_q))
 \end{equation}
 
 Explanation: The first term sums up the energy needed to read each input
 location according to the RHS side of the statement $S_q$, the middle term denotes the energy needed to compute the function ${\cal F}_q$, and the right term denotes the energy needed to write this result into the location $L(x_q) \in \mathrm T$.
 
\begin{figure}[ht]
    \centering
    \begin{tikzpicture}[
    >=Stealth,
    every node/.style={font=\small},
    ring/.style={
        draw=black!60,
        circle,
        minimum size=8mm,
        inner sep=0pt,
        line width=0.8pt
    },
    fr/.style={
        draw=black!60,
        circle,
        minimum size=11mm,
        inner sep=0pt,
        line width=0.9pt
    }
]

\definecolor{topfill}{HTML}{DCEAF7}   
\definecolor{ffill}{HTML}{E6E6E6}     
\definecolor{yfill}{HTML}{F8D7DA}     

\node[ring, fill=topfill] (v1) at (-2,1.5) {};
\node                     (d1) at (-1.15,1.5) {$\cdots$};
\node[ring, fill=topfill] (xr) at (0,1.5) {$x_{q,r}^*$};
\node                     (d2) at (1.15,1.5) {$\cdots$};
\node[ring, fill=topfill] (v2) at (2,1.5) {};

\node[fr, fill=ffill]   (fq) at (0,0) {${\cal F}_q$};
\node[ring, fill=yfill] (y)  at (0,-1.4) {$x_q$};

\draw[->] (v1.south east) -- (fq.north west);

\draw[->] (xr.south) -- 
    node[midway, right=1pt] {\scriptsize $d_j,d_k$}
    (fq.north);

\draw[->] (v2.south west) -- (fq.north east);

\draw[->] (fq.south) -- (y.north);

\end{tikzpicture}
    \caption{RDG of a computational statement $S_q$ as given in Eq.~(\ref{eq:displacement_1})} 
    \label{fig:rdg_eq5}
    \end{figure}
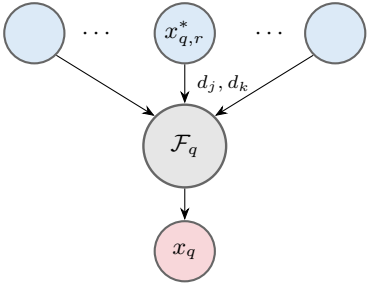

To systematically capture these dependencies, our analysis tool 
uses an internal representation called {\em reduced dependence graph (RDG)}, a directed multigraph where nodes represent inputs, outputs, and computations, and edges capture data dependencies between them over the iteration space. The RDG of a computational statement
after tiling as given in Eq.~(\ref{eq:displacement_1})
can be represented as shown in Figure~\ref{fig:rdg_eq5}.

\noindent
\subsubsection{Energy Memory-Related Statements}
For a statement $S_q^* \in M$, we estimate the energy by statement similarly:
\begin{equation}
\label{eq:energy_location}
   E_q^M = E(L(x_{q,r})) + E(L(x_{q,r}^*))
\end{equation}
 Explanation: The first term calculates the energy needed to read the right-hand-side variable
 $x_{q,r}$ of statement $S_q^*$, and the second term
 denotes the energy needed to copy its value to location $L(x_{q,r}^*) \in \mathrm T$.
Finally, holding for both types of statements, the energy of a read or write access
of any variable $x$ depends on its location $L(x) \in {\mathrm T}$ as follows:
\begin{equation*}
\label{eq:energy_per_class}
        E(L(x)) =
        \begin{cases}
        E(\mathrm{DR})+E(\mathrm{IOb})+E(\mathrm{ID})&\text{if } x \in \{I\}\\
        E(\mathrm{DR})+E(\mathrm{IOb})+E(\mathrm{OD})&\text{elsif } x \in \{O\}\\
        E(\mathrm{RD}) \; \; \;  \text{elsif } \mathbf{d}_j = \mathbf{d}_k = {0}\\
        E(\mathrm{FD}) \quad \text{elsif } \mathbf{d}_j \neq 0 \wedge \mathbf{d}_k = {0}\\
        E(\mathrm{ID}) \; \quad \text{else } (\mathbf{d}_k \neq {0})
        \end{cases}
\end{equation*}

Explanations: The first two cases hold when $x$ is an input variable and an output variable.
Here, the energy accounts for fetching the variable from DRAM into an I/O buffer and from there to an input/output register of a PE. 

\begin{example}
In the GESUMMV kernel introduced in Example~\ref{ex_1}, $A$, $B$, and $X$ are the names of \emph{input variables} (appearing in statements $S_1, S_3$ and $S_4$. Therefore, the execution of any statement involving an instance of such tensors requires a transport of the related variable instance from DRAM to an I/O buffer, and from there to an input register of a PE as indicated by the incoming green arrows at the left and top of the PE grid in Figure~\ref{fig:tcpa_arch_mapping}. Similarly, variable $Y$ in statement $S_{11}$ is an \emph{output variable}; each instance of it is written to an output register and then over an I/O buffer back to the DRAM, as illustrated by the outgoing green arrows at the right of the PE grid in Figure~\ref{fig:tcpa_arch_mapping}. 
\end{example} 

Otherwise, the location $L$ of a variable $x$ is a register ($\mathrm {RD}$), if the data is stored tile-locally (zero dependence vectors). 

\begin{example}
In the GESUMMV example, statements $S_5$ and $S_8$ are examples of local register accesses ($\mathrm {RD}$).
These PE-local data accesses are illustrated by circular blue arrows within each iteration node in Figure~\ref{fig:tcpa_arch_mapping}.
\end{example}

Otherwise, if only the intra-tile dependence vector $\mathbf{d}_j$ is non-zero, the variable might get used again in the very same PE later and is thus stored in an $\mathrm{FD}$ register.

\begin{example}
In the GESUMMV example, statements $S_2$, $S_7$, and $S_{10}$ are examples of transport statements
that, after tiling, produce a statement in which the left-hand-side variable is indexed by $(j_0,j_1,k_0,k_1)$ 
and is assigned the value of a right-hand-side variable indexed by $(j_0,j_1-1,k_0,k_1)$,  $\mathbf{d}_j \neq 0 \wedge \mathbf{d}_k = {0}$ holds, and the RHS variable will thus be stored in an $\mathrm{FD}$ register for PE-internal re-use.
These tile-local data dependencies are illustrated by the yellow arrows in Figure~\ref{fig:tcpa_arch_mapping}.
\end{example}

Else, the data comes from a different tile and thus PE and must be read into an $\mathrm{ID}$ register.

\begin{example}
For the given GESUMMV example, inter-tile dependencies may arise due to tiling, see, e.g., the
statement $S_7^*$ described in Example~\ref{ex_2} is split into two statements, of which the second has the dependence vector $\mathbf{d}^{\ast} = (0,1-p_1,0,1)^{\textrm{T}}$.
This dependence represents an inter-tile communication along the $k_1$-direction. For an iteration vector $(j_0,j_1,k_0,k_1)^{\textrm{T}}$, the data thus comes from iteration $(j_0,j_1,k_0,k_1-1)$, corresponding to the left neighbor tile (PE). The data is therefore stored in an $\mathrm{ID}$ register, as is illustrated by the orange arrows in Figure~\ref{fig:tcpa_arch_mapping}.
\end{example}
    
\subsection{Total Energy Evaluation}
With the above analysis of energy estimates per statement and iteration, we can finally derive estimates of the total energy $E_{\mathrm{tot}}$ for execution of a parametric loop nest by summing up the energies per statement after tiling multiplied by the volume ${ S}_q$ (number of integer points of) the polyhedral space where statement $S_q$ is defined:
\begin{equation}\label{eq:total_energy}
E_{\mathrm{tot}} = \sum_{S_q \in C} \operatorname{Vol}({ S}_q) \cdot E_q^C + \sum_{S_q \in M} \operatorname{Vol}({ S}_q) \cdot E_q^M.
\end{equation}

\subsection{Symbolic Volume Computation}
\label{sec:symbolic_volume}
        In the previous section, we provided a generic energy analysis per statement of a loop
        nest and a formula for the total energy for executing a given loop nest by
        multiplying these estimates by the number of integer points in the corresponding parametric polyhedra.
        For computational statements $S_q \in C$ as described in Eq.~(\ref{eq:displacement_1}), the number of iterations after tiling is obtained from the tiled iteration space.
            \begin{equation}
            \operatorname{Vol}({ S}_q)
            =
            |\big\{\mathbf{i}\mid
            \mathbf{i} = \mathbf{j} + P\mathbf{k} \land \mathbf{i} \in \mathcal{I}_q \cap \mathcal{I}
            \big\}|
            \label{eq:vol_of_computation}
            \end{equation}

        For memory-related statements of type $S_q^* \in M$ as described in Eq.~(\ref{eq:displacement_2}), the execution count of the associated memory accesses additionally depends on the dependencies $\mathbf{d}_j$ and $\mathbf{d}_k$. Therefore, for $S_q \in T$, the volume is given by
            \begin{equation}
            \operatorname{Vol}(S_q^{\ast})
            =
            |\big\{
            \mathbf{i}\mid
            \mathbf{i} =
            \mathbf{j} + P\mathbf{k}
            \land
            \mathbf{j} - \mathbf{d}_j \in \mathcal{J}
            \land
            \mathbf{i} \in
            \mathcal{I}_q \cap \mathcal{I} 
            \big\}|
            \label{eq:vol_of_data_transfer}
            \end{equation}

            The computation of such volumes corresponds to counting the number of integer points in parametric polyhedra. This can be performed symbolically using Barvinok's algorithm~\cite{barvinok} as implemented in the Integer Set Library~(ISL)~\cite{isl}. The resulting volumes are returned as {\em piecewise quasi-polynomial functions} of the loop bounds $N_i$ and tile size parameters $p_i$, which can then be inserted into Eq.~(\ref{eq:total_energy}) to determine
            the total energy consumption $E_{\mathrm{tot}}$ of all statements as executed in a given loop nest.

        \begin{example}
        To illustrate the symbolic volume computation, we consider statement $S_{7}^{\ast}$ of the GESUMMV kernel introduced in Example~\ref{ex_2}. 
        The transformed statement is decomposed into two cases based on the location of the source operand.
        The iteration spaces of $S_7^{\ast 1}$ and $S_7^{\ast 2}$ are given as follows
        with $t_i = \lceil N_i / p_i \rceil$:
        \begin{equation*}
        \mathcal{I}_{7}^{\ast 1}
        =
        \left\{
        (j_0,j_1,k_0,k_1)\in\mathbb{Z}^4
        \;\middle|\;
        \begin{aligned}
        &0 \le j_0 < p_0 \land 0 \le j_1 < p_1 \land\\
        &0 \le k_0 < t_0 \land 0 \le k_1 < t_1 \land\\
        &0 \le j_0 + p_0 k_0 < N_0\land\\
        &0 < j_1 + p_1 k_1 < N_1\land\\
        &0 \le j_1 - 1 < p_1
        \end{aligned}
        \right\}
        \end{equation*}
        \begin{equation*}
        \mathcal{I}_{7}^{\ast 2}
        =
        \left\{
        (j_0,j_1,k_0,k_1)\in\mathbb{Z}^4
        \;\middle|\;
        \begin{aligned}
        &0 \le j_0 < p_0 \land 0 \le j_1 < p_1 \land\\
        &0 \le k_0 < t_0 \land 0 \le k_1 < t_1 \land\\
        &0 \le j_0 + p_0 k_0 < N_0 \land\\
        &0 < j_1 + p_1 k_1 < N_1 \land\\
        &0 \le j_1 + p_1 - 1 < p_1
        \end{aligned}
        \right\}
        \end{equation*}

        
        Note that the above inequalities do contain non-linear terms like the expression $... \leq j + p \cdot k < ...$ (i.e., the product of the parameters $p$ and $k$). 
        But for a given (fixed) processor array size, 
        $k$ (the processor element index in a given processor array dimension with a total of $t$ elements in this dimension) is bounded, i.e., $0 \leq k < t$. Therefore, we can unfold the respective inequality constraints practically
        as follows: $... \leq \{j + p \cdot 0,\quad j+ p \cdot 1,\quad \dots,\quad j+ p (t-1)\} < ...$.\footnote{Interestingly, the symbolic energy analysis time remains on the order of 1 minute, even for large processor arrays of size $50 \times 50 = \num{2500}$ processors.} 
        By applying Barvinok's algorithm \cite{barvinok1994polynomial}, the integer-point count for $S_{7}^{\ast 1}$ and $S_{7}^{\ast 2} \in M$ is then obtained for the example of a $2 \times 2$ processor array target as shown in 
        Figure~\ref{fig:tcpa_arch_mapping}:
        \begin{equation*}
            \operatorname{vol}({{S}_{7}^{\ast 1}})
            =
            |{\mathcal{I}_{7}^{\ast 1}}|
            =
            \begin{cases}
            4\,p_0(p_1-1) & 
            \begin{aligned}[t]
            \text{if }  &0 < p_0 \land 2p_0 < N_0 \land\\
            &p_1 \ge 2 \land 2p_1 < N_1
            \end{aligned}
            \\
            2\,N_0(p_1-1) & 
            \begin{aligned}[t]
            \text{if } & N_0 > 0 \land 2p_0 \ge N_0\land\\
            &p_1 \ge 2 \land 2p_1 < N_1
            \end{aligned}
            \\
            (2N_1-4)\,p_0 & 
            \begin{aligned}[t]
            \text{if } &0 < p_0 \land 2p_0 < N_0\land\\
            &p_1 \le N_1-2 \land\\
            &2p_1 \ge N_1
            \end{aligned}
            \\
            N_0(N_1-2) & 
            \begin{aligned}[t]
            \text{if } & N_0 > 0 \land 2p_0 \ge N_0\land\\
            &p_1 \le N_1-2 \land\\
            &2p_1 \ge N_1
            \end{aligned}
            \\
            0 & \text{otherwise}
            \end{cases}
            \end{equation*}
        The above expression corresponds to the intra-tile case, where the dependence is resolved locally within a tile.
            \begin{equation*}
            \operatorname{vol}({{S}_{7}^{\ast 2}})
            =
            |{\mathcal{I}_{7}^{\ast 2}}|
            =
            \begin{cases}
            2\,p_0 &\text{if}  \ 0 < p_0 < {N_0}/{2} \land 0 < p_1 < N_1\\
            N_0 &  
            \begin{aligned}[t]
            &\text{if} \  N_0 > 0 \land p_0 \ge {N_0}/{2} \land\\
            & 0 < p_1 < N_1
            \end{aligned}\\
            0 & \text{otherwise.}
            \end{cases}
            \end{equation*}
        
        This expression corresponds to the inter-tile case, where the dependence crosses tile boundaries. The volume remains fully parametric in the loop bounds $N_i$ and tile sizes $p_i$, and thus can be evaluated very simply by just inserting concrete values for the parameters $N_0, N_1, p_0,$ and $p_1$. E.g., for the values exemplified in Figure~\ref{fig:tcpa_arch_mapping} with a shown iteration space of size $N_0 \times N_1 = 4 \times 5 = 20$ and tiles of size $2 \times 3$, the resulting volumes are ${\operatorname{Vol}}(\mathcal{S}_{7}^{\ast 1}) = 12$ and ${\operatorname{Vol}}(\mathcal{S}_{7}^{\ast 2}) = 4$, corresponding to 12 intra-tile dependences (exactly matches the number of yellow arrows in $j_1$ direction) and 4 inter-tile dependences (exactly matches the number of orange arrows in $k_1$ direction).
        
        Using Eq.~(\ref{eq:energy_location}), the energy of $S_{7}^{\ast 1}$ is given by one FD read and one RD write,
        \[
        E_{7^{\ast 1}}^{M}=E(\mathrm{FD})+E(\mathrm{RD})=0.35+0.12=\qty{0.47}{\pico\joule}.
        \]
        Similarly, the energy of $S_{7}^{\ast 2}$ is given by one ID read and one RD write,
        \[
        E_{7^{\ast 2}}^{M}=E(\mathrm{ID})+E(\mathrm{RD})=0.24+0.12=\qty{0.36}{\pico\joule}.
        \]
        For the concrete configuration, the contribution of these two statements to $E_{\mathrm{tot}}$ in Eq.~(\ref{eq:total_energy}) evaluates to
        \begin{equation*}
        \operatorname{Vol}(\mathcal{S}_{7}^{\ast 1})E_{7^{\ast 1}}^{M}
        +
        \operatorname{Vol}(\mathcal{S}_{7}^{\ast 2})E_{7^{\ast 2}}^{M}
        =
        12 \cdot 0.47 + 4 \cdot 0.36 = \qty{7.08}{\pico\joule}.
        \end{equation*}
        
    This energy corresponds to the total energy contribution of statement $S_{7}^{\ast}$ for the given configuration. The overall kernel energy is obtained by summing the contributions of all statements according to Eq.~(\ref{eq:total_energy}).
    \end{example}

\section{Experimental results}\label{sec:Experimental_results}
    This section evaluates the proposed symbolic energy analysis framework from two complementary perspectives. First, we validate the accuracy and analysis-time efficiency of the symbolic approach by deriving parametric volumes for both computations and memory operations, and comparing these with counts obtained from a cycle-accurate simulation of the execution of a loop nest. Second, we analyze energy and latency scaling with increasing problem sizes and show that the symbolic model provides a scalable basis for energy evaluation, making it particularly suitable
    to explore application-specific architecture sizing.
    
\begin{figure}[!t]
    \centering
    \includegraphics[width=0.96\columnwidth,trim=8 12 8 10,clip]{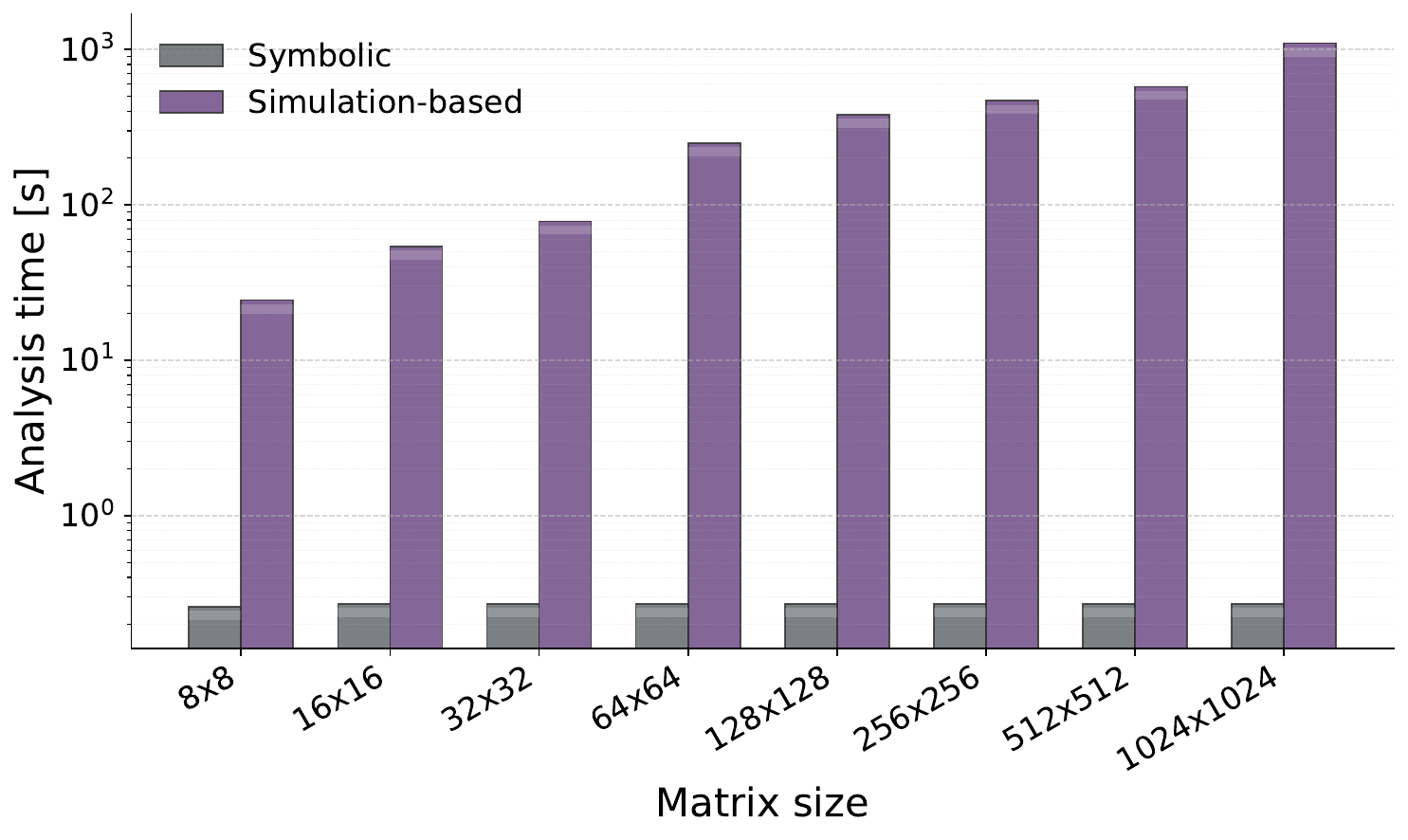}
    \caption{Comparison of symbolic and simulation-based analysis times for the GESUMMV benchmark on an 8×8 PE array across increasing matrix sizes. The symbolic approach remains nearly constant, while simulation time grows rapidly with problem size.}
    \label{fig:sim_vs_analytical_exe_time}
    \vspace{-6pt}
\end{figure}
        
    \subsection{Validation of Symbolic Volume Computation Results}
    \label{subsec:validation}
        We validate the accuracy of the symbolic volume computation results by comparing analytically derived integer-point counts for both data transfer and computation against reference counts obtained from a cycle-accurate simulator. The simulator operates using an XML-based architectural description that captures the entire TCPA architecture. Based on this description, the TCPA compiler~\cite{WitteraufWHT21} automatically tiles the input loop programs, maps the tiles onto the processor array, and schedules their execution. During simulation, all data transfers and computations are tracked, yielding exact reference counts. Each selected PolyBench kernel~\cite{PolyBench} is then simulated for a specified architectural configuration to obtain ground-truth memory access and computation counts.
        
        Note that our analytical method can analyze the energy of a loop kernel symbolically without requiring concrete loop bounds. The resulting volumes are expressed as closed-form quasi-polynomials. In the following, we evaluated our approach across eight different benchmarks from~\cite{PolyBench} for varying problem sizes and architectural parameters. The analytically derived access 
        counts and obtained total energy values match the simulation results exactly, confirming the accuracy of the symbolic formulation.
        
        Figure~\ref{fig:sim_vs_analytical_exe_time} compares the analysis time of simulation-based counting and the proposed symbolic method for the GESUMMV benchmark mapped onto an $8\times8$ PE array. The simulation-based approach exhibits a rapid increase in analysis time with increasing matrix size, as it explicitly executes all loop iterations, instruction activities, and memory accesses within the cycle-accurate model. As the iteration space grows quadratically with the matrix dimension, the simulation cost scales accordingly.
        
        In contrast, the symbolic approach evaluates one-time calculated closed-form expressions for computation and memory access statements, independent of the number of loop iterations. As a result, the analysis time remains almost constant at less than $0.5 \ \unit{s}$  across all evaluated problem sizes. This decoupling of analysis cost from the dynamic execution makes the symbolic method scalable and suitable for exploring large loop bounds that are impractical for simulation-based analysis.

        \begin{figure}[t]
            \centering
            \includegraphics[width=0.95\columnwidth,trim=8 12 8 10,clip]{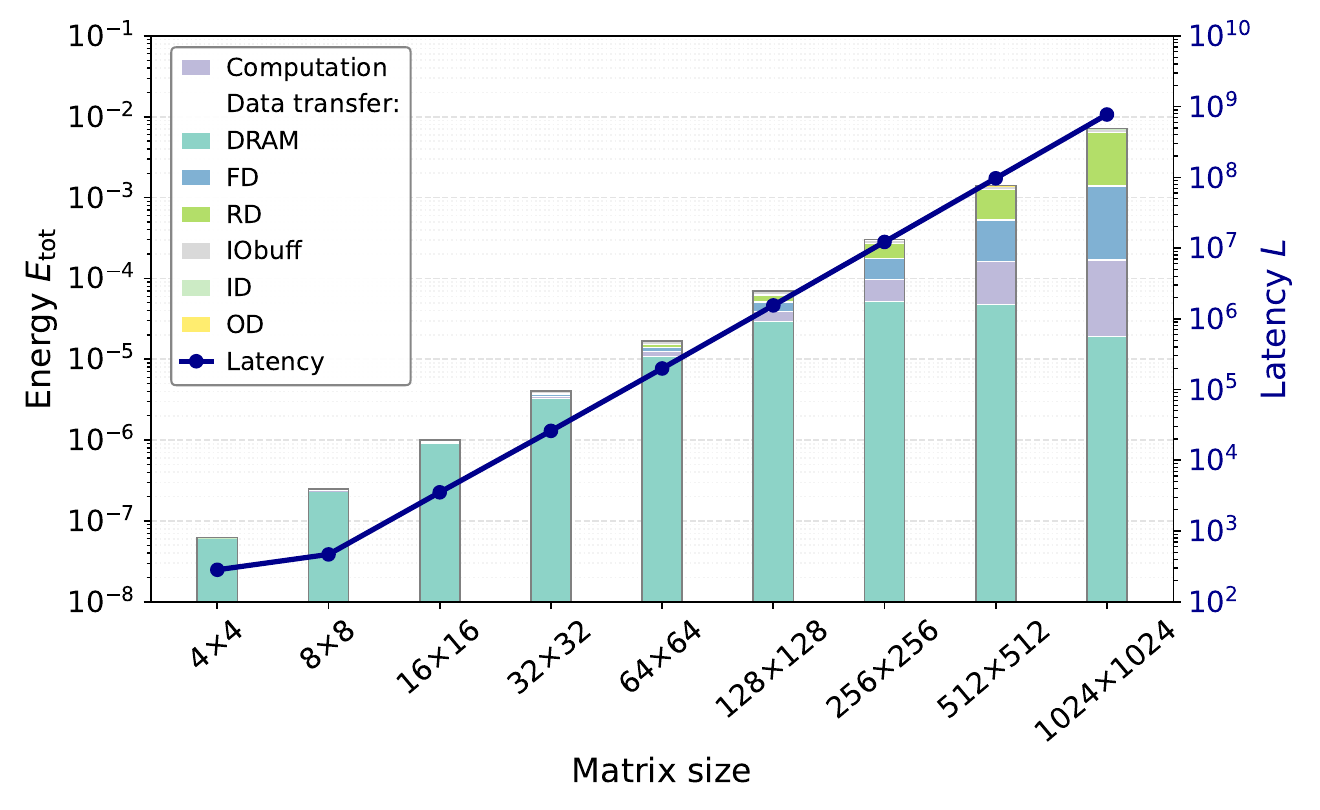}
            \caption{Energy $E_{\mathrm{tot}}$ and latency $L$ vs.\ matrix size for GEMM on an 8×8 PE grid TCPA, showing a shift from DRAM-dominated energy to increasing on-chip communication (FD/RD) at larger scales due to growing tile sizes.}
            \label{fig:mm-energy-vs-bound}
            \vspace{-6pt}
        \end{figure}
        


\subsection{Energy and Latency Scaling Analysis}
Since the proposed analysis derives expressions, it enables a fast evaluation of bounds and exploration of architectural configurations.
In the following, we study the total energy $E_{\mathrm{tot}}$ and latency $L$ of a given loop nest with increasing loop bounds. Thereby, we provide a fine-grained breakdown of energy contributions across different access locations $L(x) \in \mathrm{T}$, revealing how memory accesses and computations contribute individually. As will be shown, computation energy remains relatively small compared to data transfer energy across all configurations. The total energy and latency are computed using Eq.~\eqref{eq:total_energy} and~\eqref{eq:global-latency}, respectively. For this analysis, we consider the GEMM kernel~\cite{PolyBench}, where the iteration space grows cubically with problem size (i.e., $O(N^3)$). As loop bounds increase, both computation and data movement grow rapidly. Figure~\ref{fig:mm-energy-vs-bound} illustrates how total energy and latency increase with matrix size for GEMM, analytically evaluated for an $8\times8$ PE grid. As expected, both metrics grow rapidly with increasing loop bounds due to the cubic growth of the iteration space.
        
        The energy breakdown further reveals how different components scale. For smaller problem sizes, DRAM accesses dominate the total energy consumption. However, as the loop bounds increase, the relative contribution of DRAM energy decreases, while the energy associated with on-chip storage and communication—such as FD and RD registers—as well as computation, increases. This shift is primarily due to larger tile sizes, which increase intra-tile data reuse and consequently amplify activity within local storage locations. This growth is not strictly proportional to the loop bounds, as data accessed from different locations scales differently depending on tile sizes and data dependencies.
In many existing energy estimation approaches, activity counts are obtained by simulating a fixed workload instance for smaller loop bounds and extrapolating to larger ones, since simulation is not efficient at handling larger bounds.
        Our proposed symbolic analysis does not suffer from scalability.
        
Since not only the energy estimates are parametric but also the schedules, performance metrics such as latency, throughput, and thus also energy efficiency can be computed analytically.
This paves the way for a rapid comparison of architectural configurations and supports DSE to identify suitable accelerator architectures for a huge number of loop applications.
        
\section{Conclusion}\label{sec:Conclusion}
    This paper introduces a symbolic methodology for energy analysis of loop nests when mapped and scheduled on parallel processor array accelerator architectures. In contrast to simulation-based approaches, which require an explicit execution of all loop iterations, all memory accesses, and all operations, our method derives closed-form expressions for computations and memory accesses directly from the program representation and mapping. By combining symbolic volume computation of polyhedral spaces with pre-characterized energy margins per access and operation types, the approach enables a symbolic and accurate estimation of energy without requiring time-intensive cycle-accurate simulations for each setting of loop bounds.
    
    The proposed analysis lifts energy evaluation from a simulation-based process to a polyhedral-model-based formulation that is applicable at early design stages. Once the symbolic expressions are derived, the total energy can be evaluated efficiently for different loop bounds and architecture configurations without repeated analysis. This is particularly important for loop-intensive applications, where simulation time increases rapidly with problem size, while the symbolic evaluation remains nearly constant, providing a scalable basis for design space exploration.
    
    Our evaluations have demonstrated that our symbolic and cycle-accurate simulation-based energy analysis approaches match in their results, with the symbolic approach also providing a fine-grained view of the contributions of different types of memory and register accesses. This makes it possible to not only study the total energy, but also the influence of array size, mapping, scheduling, and data movement across different access locations.
    
Overall, the presented framework provides a practical and efficient approach to energy estimation for processor-array accelerators.
By enabling fast symbolic evaluation of energy, latency, and related performance metrics from parametric loop bounds, it supports a fast analysis of architectural configurations and helps to identify suitable accelerator architectures for a given application.

{\renewcommand*{\bibfont}{\footnotesize}
\printbibliography}

@inproceedings{TCPA1,
  author       = {Dmitrij Kissler and
                  Frank Hannig and
                  Alexey Kupriyanov and
                  J{\"{u}}rgen Teich},
  title        = {A highly parameterizable parallel processor array architecture},
  booktitle    = {{IEEE} Int. Conf. on Field Programmable Technology (FPT), Bangkok, Thailand},
  pages        = {105--112},
  publisher    = {IEEE},
  year         = {2006},
  doi          = {10.1109/FPT.2006.270293},
}

@inproceedings{alpaka,
  author       = {Dominik Walter and
                  Marcel Brand and
                  Christian Heidorn and
                  Michael Witterauf and
                  Frank Hannig and
                  J{\"{u}}rgen Teich},
  title        = {{ALPACA:} An Accelerator Chip for Nested Loop Programs},
  booktitle    = {Int. Symposium on Circuits and Systems (ISCAS)},
  pages        = {1--5},
  publisher    = {{IEEE}},
  year         = {2024},
  doi          = {10.1109/ISCAS58744.2024.10558549},
}

@phdthesis{Teich93,
  author       = {J{\"{u}}rgen Teich},
  title        = {A compiler for application specific processor arrays},
  school       = {Saarland University, Germany},
  year         = {1993},
  url          = {https://d-nb.info/931645875},
  isbn         = {978-3-86111-701-8},
  bibsource    = {dblp computer science bibliography, https://dblp.org}
}

@article{energy_weights,
  author       = {Ardavan Pedram and
                  Stephen Richardson and
                  Mark Horowitz and
                  Sameh Galal and
                  Shahar Kvatinsky},
  title        = {Dark Memory and Accelerator-Rich System Optimization in the Dark Silicon Era},
  journal      = {{IEEE} Des. Test},
  volume       = {34},
  number       = {2},
  pages        = {39--50},
  year         = {2017},
  doi          = {10.1109/MDAT.2016.2573586},
}

@article{TeichTH4,
  author       = {J{\"{u}}rgen Teich and
                  Alexandru Tanase and
                  Frank Hannig},
  title        = {Symbolic Mapping of Loop Programs onto Processor Arrays},
  journal      = {J. Signal Process. Syst.},
  volume       = {77},
  number       = {1-2},
  pages        = {31--59},
  year         = {2014},
  doi          = {10.1007/S11265-014-0905-0},
}

@article{WitteraufWHT21,
  author  = {Witterauf, M. and Walter, D. and Hannig, F. and Teich, J.},
  title   = {Symbolic Loop Compilation for {T}ightly {C}oupled {P}rocessor {A}rrays},
  journal = {ACM Trans. Embedded Comput. Syst. (TECS)},
  volume  = {20},
  number  = {5},
  year    = {2021},
  pages   = {1--31}
}

@article{TeichT93,
  author       = {J{\"{u}}rgen Teich and
                  Lothar Thiele},
  title        = {Partitioning of processor arrays: a piecewise regular approach},
  journal      = {Integr.},
  volume       = {14},
  number       = {3},
  pages        = {297--332},
  year         = {1993},
  doi          = {10.1016/0167-9260(93)90013-3},
}

@article{TeichTZ97,
  author       = {J{\"{u}}rgen Teich and
                  Lothar Thiele and
                  Lee Z. Zhang},
  title        = {Partitioning Processor Arrays under Resource Constraints},
  journal      = {J. {VLSI} Signal Process.},
  volume       = {17},
  number       = {1},
  pages        = {5--20},
  year         = {1997},
  doi          = {10.1023/A:1007935215591},
}

@misc{PolyBench,
  author       = {Pouchet, Louis-No{\"e}l},
  title        = {PolyBench: The Polyhedral Benchmark Suite},
  note         = {Accessed: Oct. 10, 2025}
}

@inproceedings{AccelWattch,
  author       = {Vijay Kandiah and
                  Scott Peverelle and
                  Mahmoud Khairy and
                  Junrui Pan and
                  Amogh Manjunath and
                  Timothy G. Rogers and
                  Tor M. Aamodt and
                  Nikos Hardavellas},
  title        = {AccelWattch: {A} Power Modeling Framework for Modern GPUs},
  booktitle    = {Int. Symposium on Microarchitecture (MICRO)},
  pages        = {738--753},
  publisher    = {{ACM}},
  year         = {2021},
  doi          = {10.1145/3466752.3480063},
}

@article{CGRA-EAM,
  author       = {Mark Wijtvliet and
                  Henk Corporaal and
                  Akash Kumar},
  title        = {{CGRA-EAM} - Rapid Energy and Area Estimation for Coarse-grained Reconfigurable
                  Architectures},
  journal      = {{ACM} Trans. Reconfigurable Technol. Syst.},
  volume       = {14},
  number       = {4},
  pages        = {19:1--19:28},
  year         = {2021},
  doi          = {10.1145/3468874},
}

@inproceedings{Accelergy,
  author       = {Yannan Nellie Wu and
                  Joel S. Emer and
                  Vivienne Sze},
  editor       = {David Z. Pan},
  title        = {Accelergy: An Architecture-Level Energy Estimation Methodology for
                  Accelerator Designs},
  booktitle    = {Proceedings of the Int. Conf. on Computer-Aided Design,
                  {ICCAD} 2019, Westminster, CO, USA, November 4-7, 2019},
  pages        = {1--8},
  publisher    = {{ACM}},
  year         = {2019},
  doi          = {10.1109/ICCAD45719.2019.8942149},
}

@inproceedings{IO_transfor,
  author       = {Dominik Walter and
                  Michael Witterauf and
                  J{\"{u}}rgen Teich},
  title        = {Real-time Scheduling of {I/O} Transfers for Massively Parallel Processor
                  Arrays},
  booktitle    = {18th {ACM/IEEE} Int. Conf. on Formal Methods and Models
                  for System Design, {MEMOCODE} 2020, Jaipur, India, December 2-4, 2020},
  pages        = {1--11},
  publisher    = {{IEEE}},
  year         = {2020},
  doi          = {10.1109/MEMOCODE51338.2020.9315179},
}

@inproceedings{sym_tiling_sche,
  author       = {Michael Witterauf and
                  Alexandru Tanase and
                  Frank Hannig and
                  J{\"{u}}rgen Teich},
  title        = {Modulo scheduling of symbolically tiled loops for {T}ightly {C}oupled
                  {P}rocessor {A}rrays},
  booktitle    = {Int. Conf. on Application-specific Systems,
                  Architectures and Processors (ASAP)},
  pages        = {58--66},
  publisher    = {{IEEE}},
  year         = {2016},
  doi          = {10.1109/ASAP.2016.7760773},
}

@inproceedings{isl,
  author       = {Sven Verdoolaege},
  editor       = {Komei Fukuda and
                  Joris van der Hoeven and
                  Michael Joswig and
                  Nobuki Takayama},
  title        = {\emph{isl}: An Integer Set Library for the Polyhedral Model},
  booktitle    = {Third Int. Congress on Mathematical Software (ICMS)},
  volume       = {6327},
  pages        = {299--302},
  publisher    = {Springer},
  year         = {2010},
  doi          = {10.1007/978-3-642-15582-6_49},
}

@book{barvinok,
  author       = {Alexander I. Barvinok},
  title        = {A course in convexity},
  series       = {Graduate studies in mathematics},
  volume       = {54},
  publisher    = {American Mathematical Society},
  year         = {2002},
  isbn         = {978-0-8218-2968-4},
}

@article{Eyeriss,
  author       = {Yu{-}Hsin Chen and
                  Tushar Krishna and
                  Joel S. Emer and
                  Vivienne Sze},
  title        = {Eyeriss: An Energy-Efficient Reconfigurable Accelerator for Deep Convolutional
                  Neural Networks},
  journal      = {{IEEE} J. Solid State Circuits},
  volume       = {52},
  number       = {1},
  pages        = {127--138},
  year         = {2017},
  doi          = {10.1109/JSSC.2016.2616357},
}

@inproceedings{Timeloop,
  author       = {Angshuman Parashar and
                  Priyanka Raina and
                  Yakun Sophia Shao and
                  Yu{-}Hsin Chen and
                  Victor A. Ying and
                  Anurag Mukkara and
                  Rangharajan Venkatesan and
                  Brucek Khailany and
                  Stephen W. Keckler and
                  Joel S. Emer},
  title        = {Timeloop: {A} Systematic Approach to {DNN} Accelerator Evaluation},
  booktitle    = {{IEEE} Int. Symposium on Performance Analysis of Systems
                  and Software, {ISPASS}, Madison, USA},
  pages        = {304--315},
  publisher    = {{IEEE}},
  year         = {2019},
  doi          = {10.1109/ISPASS.2019.00042},
}

@inproceedings{Verdoolaege04,
author = {Verdoolaege, Sven and Seghir, Rachid and Beyls, Kristof and Loechner, Vincent and Bruynooghe, Maurice},
title = {Analytical computation of Ehrhart polynomials: enabling more compiler analyses and optimizations},
year = {2004},
isbn = {1581138903},
publisher = {ACM},
url = {https://doi.org/10.1145/1023833.1023868},
doi = {10.1145/1023833.1023868},
booktitle = {Int. Conf. on Compilers, Architecture, and Synthesis for Embedded Systems (CASES)},
pages = {248–258},
numpages = {11},
}

@article{TECS14,
  author    = {Frank Hannig and
               Vahid Lari and
               Srinivas Boppu and
               Alexandru Tanase and
               Oliver Reiche},
  title     = {Invasive Tightly-Coupled Processor Arrays: {A} Domain-Specific Architecture/Compiler
               Co-Design Approach},
  journal   = {{ACM} Trans. Embedded Comput. Syst.},
  volume    = {13},
  number    = {4s},
  pages     = {133:1--133:29},
  year      = {2014},
  url       = {https://doi.org/10.1145/2584660},
  doi       = {10.1145/2584660},
  bibsource = {dblp computer science bibliography, https://dblp.org}
}

@article{barvinok1994polynomial,
  title={A polynomial time algorithm for counting integral points in polyhedra when the dimension is fixed},
  author={Barvinok, Alexander I.},
  journal={Mathematics of Operations Research},
  volume={19},
  number={4},
  pages={769--779},
  year={1994}
}

@inproceedings{cache_analysis_using_barvinoks,
  title={Analytical computation of Ehrhart polynomials and its application in compile-time generated cache hints},
  author={Seghir, Rachid and Verdoolaege, Sven and Beyls, Kristof and Loechner, Vincent},
  booktitle={Int. Conf. on Compilers, Architecture and Synthesis for Embedded Systems (CASES)},
  year={2004}
}

@inproceedings{26_CGRA_Overview,
    author = {Wijtvliet, Mark and Waeijen, Luc and Corporaal, Henk},
    title = {Coarse grained reconfigurable architectures in the past 25 years: Overview and classification},
    booktitle = {SAMOS},
    year = {2016},
    pages = {235-244},
    doi = {10.1109/SAMOS.2016.7818353},
}

@article{24_CGRA_Taxonomy,
    author = {Liu, Leibo and Zhu, Jianfeng and Li, Zhaoshi and Lu, Yanan and Deng, Yangdong and Han, Jie and Yin, Shouyi and Wei, Shaojun},
    title = {A Survey of Coarse-Grained Reconfigurable Architecture and Design: Taxonomy, Challenges, and Applications},
    journal = {Comput. Surv.},
    year = {2019},
    volume = {52},
    number = {6},
    pages = {118:1--118:39},
    doi = {10.1145/3357375},
}

@inproceedings{cgra_comp,
    author = {Walter, Dominik and Halm, Marita and Seidel, Daniel and Ghosh, Indrayudh and Heidorn, Christian and Hannig, Frank and Teich, Jürgen},
    booktitle = {29. Workshop zu Methoden und Beschreibungssprachen zur Modellierung und Verifikation von Schaltungen und Systemen},
    date = {2026},
    title = {{Modeling} and {Mapping} of {Regular} {Nested} {Loops} on {Processor} {Arrays}: {CGRAs} vs. {TCPAs}},
    venue = {Würzburg},
    year = {2026}
}
\end{document}